\documentclass[groupedaddress,aps,pra,superscriptaddress,showpacs,twocolumn,prl]{revtex4}%
\usepackage{epsfig,dsfont,amssymb,amsmath,amsthm,amsfonts,amsbsy,mathrsfs}
\usepackage{graphicx, color}
\usepackage{epstopdf}
\usepackage{mathdots}
\usepackage{float}
\begin{document}

\title{Polygamy relations of multipartite entanglement beyond qubits}

\author{Zhi-Xiang Jin}
\thanks{Corresponding author: jzxjinzhixiang@126.com}
\affiliation{School of Mathematical Sciences,  Capital Normal University,  Beijing 100048,  China}
\author{Shao-Ming Fei}
\thanks{Corresponding author: feishm@mail.cnu.edu.cn}
\affiliation{School of Mathematical Sciences,  Capital Normal University,  Beijing 100048,  China}
\affiliation{Max-Planck-Institute for Mathematics in the Sciences, Leipzig 04103, Germany}

\begin{abstract}

We investigate the polygamy relations related to the concurrence of assistance for any multipartite pure states. General polygamy inequalities given by the $\alpha$th $(0\leq \alpha\leq 2)$ power of concurrence of assistance is first presented for multipartite pure states in arbitrary-dimensional quantum systems. We further show that the general polygamy inequalities can even be improved to be tighter inequalities under certain conditions on the assisted entanglement of bipartite subsystems. Based on the improved polygamy relations, lower bound for distribution of bipartite entanglement is provided in a multipartite system. Moreover, the $\beta$th ($0\leq \beta \leq 1$) power of polygamy inequalities are obtained for the entanglement of assistance as a by-product, which are shown to be tighter than the existing ones. A detailed example is presented.

\end{abstract}
\maketitle

\section{INTRODUCTION}
Quantum entanglement \cite{MAN,RPMK,FMA,KSS,HPB,HPBO,JIV,CYSG} has been extensively studied due to its importance in quantum communication and quantum information processing in recent years. The study of quantum entanglement from various viewpoints has been a very active area and has led to many impressive results. Monogamy of entanglement is one of the nonintuitive phenomena of quantum physics that distinguish quantum from classical physics. Different from the classical world, it is not possible to prepare three qubits in a way that any two qubits are maximally entangled. Qualitatively, monogamy of entanglement measures the shareability of entanglement in a composite quantum system. Moreover, the monogamy property has emerged as the ingredient in the security analysis of quantum key distribution \cite{MP}.	

The monogamy relation was first quantified by Coffman, Kundu, and Wootters \cite{MK} for three qubits, which satisfies $\mathcal{E}_{A|BC}\geq \mathcal{E}_{AB} +\mathcal{E}_{AC}$. The CKW inequality shows the mutually exclusive nature of multipartite quantum entanglement in a quantitative way: more entanglement shared between two qubits ($A$ and $B$) necessarily implies less entanglement between the other two qubits ($A$ and $C$). CKW inequality was generalized for multiqubit systems \cite{TJ} and also studied intensively in more general settings \cite{gy1,gy2}. However, the CKW inequality fails for higher-dimensional quantum systems.
It is also not generally true for three-qubit systems with other entanglement measures like entanglement of formation \cite{wootters}. Monogamy of multiqubit entanglement and some higher-dimensional quantum systems were later characterized in terms of various entanglement measures \cite{ZXN,JZX,jll}.

Whereas the monogamy of entanglement shows the restricted sharability of multipartite quantum entanglement,
the distribution of entanglement in multipartite quantum systems was shown to have a dually monogamous property. Using concurrence of assistance \cite{tfs} as the measure of distributed entanglement, the polygamy of entanglement
provides a lower bound for distribution of bipartite entanglement in a multipartite system \cite{bcs}.
Polygamy of entanglement is characterized as a polygamy inequality, ${E_a}_{A|BC}\leq {E_a}_{AB} +{E_a}_{AC}$ for a tripartite quantum state $\rho_{ABC}$, where ${E_a}_{A|BC}$ is the assisted entanglement \cite{gg}
between $A$ and $BC$. Polygamy of entanglement was generalized to multiqubit systems \cite{bcs} and arbitrary dimensional multipartite states \cite{062328,295303,bcs,042332}.

The study of quantum entanglement in higher-dimensional quantum systems is of importance in quantum information processing. Monogamy and polygamy of entanglement can restrict the possible correlations between the authorized users and the eavesdroppers, which tightens security bounds in quantum cryptography. And optimized monogamy and polygamy relations give rise to finer characterizations of the entanglement distributions. Furthermore, to optimize the efficiency of entanglement usage as a resource in quantum cryptography, higher-dimensional quantum systems rather than qubits are preferred in some physical systems for stronger security in quantum key distribution \cite{gjvw}.

In this paper, we provide a tighter polygamy inequalities for arbitrary dimensional quantum systems.
General polygamy inequalities given by the $\alpha$th $(0\leq \alpha\leq 2)$ power of concurrence of assistance are first presented for multipartite pure states in arbitrary-dimensional quantum systems. We further show that the general polygamy inequalities can even be improved to be tighter inequalities under certain conditions on the assisted entanglement of bipartite subsystems. Based on the improved polygamy relations, lower bound for distribution of bipartite entanglement is provided for multipartite systems. Moreover, the $\beta$th ($0\leq \beta \leq 1$) power of polygamy inequalities are obtained for the entanglement of assistance, which are shown to be tighter than the existing ones.

For a bipartite pure state $|\psi\rangle_{AB}\in\mathds{H}_A\otimes \mathds{H}_B$, the concurrence is given by \cite{AU,PR,SA}, $C(|\psi\rangle_{AB})=\sqrt{{2\left[1-\mathrm{Tr}(\rho_A^2)\right]}}$,
where $\rho_A=\mathrm{Tr}_B(|\psi\rangle_{AB}\langle\psi|)$.
 For mixed state, $\rho_{AB}$ is defined by the convex roof extension,
$C(\rho_{AB})=\min_{\{p_i,|\psi_i\rangle\}}\sum_ip_iC(|\psi_i\rangle)$,
where the minimum is taken over all possible pure state decompositions of $\rho_{AB}=\sum\limits_{i}p_i|\psi_i\rangle\langle\psi_i|$,
with $p_i\geq0$, $\sum\limits_{i}p_i=1$ and $|\psi_i\rangle\in \mathds{H}_A\otimes \mathds{H}_B$.

For three qubit pure state $|\psi\rangle_{ABC}$, the concurrence of assistance defined  as \cite{TFS, YCS},
\begin{eqnarray*}
C_a(|\psi\rangle_{ABC})\equiv C_a(\rho_{AB})=\max\limits_{\{p_i,|\psi_i\rangle\}}\sum_ip_iC(|\psi_i\rangle),
\end{eqnarray*}
where the maximum is taken over all possible pure state decompositions of $\rho_{AB}=\mathrm{Tr}_C(|\psi\rangle_{ABC}\langle\psi|)=\sum\limits_{i}p_i|\psi_i\rangle_{AB}\langle\psi_i|.$
For pure states $\rho_{AB}=|\psi\rangle_{AB}\langle\psi|$, one has $C(|\psi\rangle_{AB})=C_a(\rho_{AB})$.

For an $N$-qubit state $\rho_{AB_1\cdots B_{N-1}}\in \mathds{H}_A\otimes \mathds{H}_{B_1}\otimes\cdots\otimes \mathds{H}_{B_{N-1}}$,
the concurrence $C(\rho_{A|B_1\cdots B_{N-1}})$  viewed as a bipartite state under the partition
$A$ and $B_1,B_2,\cdots, B_{N-1}$, satisfies the monogamy relation \cite{TJ,YKM},
\begin{eqnarray}\label{C2}
  C^2(\rho_{A|B_1,B_2\cdots,B_{N-1}})\geq \sum_{i=1}^{N-1}C^2(\rho_{AB_i}),
\end{eqnarray}
where $\rho_{AB_i}=\mathrm{Tr}_{B_1\cdots B_{i-1}B_{i+1}\cdots B_{N-1}}(\rho_{AB_1\cdots B_{N-1}})$.
Further improved monogamy relations are presented in \cite{ZXN} and \cite{JZX}.

Different from the concurrence, the concurrence of assistance for $N$-qubit states have the form \cite{bcs},
\begin{eqnarray}\label{DCA}
 C^2_a(\rho_{A|B_1,B_2\cdots,B_{N-1}})\leq \sum_{i=1}^{N-1}{C_a^2}(\rho_{AB_i}).
\end{eqnarray}

For a bipartite arbitrary dimensional pure state $|\phi\rangle_{AB}=\sum_{i=1}^{d_1}\sum_{k=1}^{d_2}a_{ik}|ik\rangle_{AB}$ in $C^{d_1}\otimes C^{d_2}$,
the concurrence is given by \cite{jpa6777}
\begin{eqnarray}\label{beyond}
C^2(|\phi\rangle_{AB})=2(1-\mathrm{Tr}(\rho_A^2))=4\sum_{i<j}^{d_1}\sum_{k<l}^{d_2}|a_{ik}a_{jl}-a_{il}a_{jk}|^2.
\end{eqnarray}
And for a mixed state $\rho_{AB}=\sum_ip_i|\phi_i\rangle_{AB}\langle\phi_i|$, from (\ref{beyond}) its concurrence of assistance satisfies $C_a(\rho_{AB})=\max_{\{p_i,|\phi_i\rangle\}}\sum_ip_iC(|\phi_i\rangle)\leq\sum_{m=1}^{D_1}\sum_{n=1}^{D_2}(\max\sum_ip_i|\langle\phi_i|(L_A^m\otimes L_B^n)|\phi_i^\ast\rangle|)=\sum_{m=1}^{D_1}\sum_{n=1}^{D_2}C_a((\rho_{AB})_{mn}):=\tau_a(\rho_{AB})$ \cite{022302}, where $D_1=d_1(d_1-1)/2,~D_2=d_2(d_2-1)/2$, $L_A^m=P_A^m(-|i\rangle_A\langle j|+|j\rangle_A\langle i|)P_A^m$, $L_B^n=P_B^n(-|k\rangle_B\langle l|+|l\rangle_B\langle k|)P_B^n$, and $P_A^m=|i\rangle_A\langle i|+|j\rangle_A\langle j|$,
$P_B^n=|k\rangle_B\langle k|+|l\rangle_B\langle l|$ are the projections onto the subspaces spanned by the local bases
$\{|i\rangle_A, |j\rangle_A\}$ and $\{|k\rangle_B, |l\rangle_B\}$, respectively. In Ref. \cite{022302}, the authors give a general polygamy inequality for any multipartite pure
state $|\phi\rangle_{A_1\cdots A_n}\in C^{d_1}\otimes\cdots\otimes C^{d_n}$
\begin{eqnarray}\label{cb}
\tau_a^2(|\phi\rangle_{A_1|A_2\cdots A_n})\leq \sum_{i=2}^n\tau_a^2(\rho_{A_1A_i}),
\end{eqnarray}
where $\rho_{A_1A_k}$ is the reduced density matrix of $|\phi\rangle_{A_1|A_2\cdots A_n}$ associated with the subsystems $A_1A_k$, $k=2,\cdots,n$.

\section{weighted POLYGAMY RELATION for CONCURRENCE of assistance}

Polygamy of entanglement states that if a multipartite state is maximally entangled with respect to a given kind of multipartite entanglement, then it must be pure \cite{bcd}. This observation implies that all maximally entangled states are necessarily uncorrelated with any other systems. One can even propose this condition as another requisite for a good multipartite entanglement quantifier. Furthermore, it is also important to note that this polygamy holds for all kinds of entanglement, that is, whenever a system reaches a maximum amount of entanglement under any partitions, it becomes ``free'' of its environment.

Therefore, for states that do not reach the maximum amount of entanglement of assistance under any partition, the polygamy inequality of entanglement provides a lower bound for the distribution of bipartite entanglement in a multipartite system. Meanwhile, the bipartite sharability of entanglement in a multipartite system gives an upper bound of the entanglement. Tighter polygamy inequalities give rise to finer characterization of the entanglement distributions, which are tightly related to the security of quantum cryptographic protocols based on entanglement \cite{MP} (it limits the amount of correlations that an eavesdropper can have with the honest parties). In the following, we give a class of polygamy inequalities that are tighter than existing ones. First, we give the definition of Hamming weight.

For any non-negative integer $j$ and its binary expansion
\begin{eqnarray*}\label{}
j=\sum_{i=0}^{n-1}j_i2^i,
\end{eqnarray*}
with $\log_2j\leq n$ and $j_i\in \{0,~1\}$, $i=0,1,\cdots,n-1$, we can always define a unique binary vector $\vec{j}$ associated with $j$,
\begin{eqnarray}\label{wj}
\vec{j}=(j_0,j_1,\cdots,j_{n-1}).
\end{eqnarray}
For the binary vector $\vec{j}$ defined in (\ref{wj}), the Hamming weight $w_H(\vec{j})$ is defined by the number of $1's$ in $\{j_0,j_1,\cdots,j_{n-1}\}$ \cite{MAN}.

{[\bf Lemma 1]}. For any real numbers $x$ and $t$, $0\leq t \leq 1$, $0\leq x \leq 1$, we have $(1+t)^x\leq 1+(2^{x}-1)t^x$.

{\sf [Proof].} Let $f(x,y)=(1+y)^x-y^x$ with $0\leq x\leq 1,~y\geq 1$. Then $\frac{\partial f}{\partial y}=x[(1+y)^{x-1}-y^{x-1}]\leq 0$. Therefore, $f(x,y)$ is an decreasing function of $y$, i.e., $f(x,y)\leq f(x,1)=2^x-1$. Set $y=\frac{1}{t},~0<t\leq 1$, we obtain $(1+t)^x\leq 1+(2^x-1)t^x$. When $t=0$, the inequality is trivial. $\Box$

The following theorem provides states that a class of polygamy inequalities satisfied by the $\alpha$-power of $\tau_a$.
For convenience, we denote ${\tau_a}(\rho_{AB_i})={\tau_a}_{AB_i}$ the concurrence of assistance $\rho_{AB_i}$ and ${\tau_a}(\rho_{A|B_0B_1\cdots B_{N-1}})={\tau_a}_{A|B_0B_1\cdots B_{N-1}}$.

{[\bf Theorem 1]}. For any multiparty pure state $\rho_{AB_0\cdots B_{N-1}}$, we have
\begin{eqnarray}\label{th12}
{\tau_a^\alpha}_{A|B_0B_1\cdots B_{N-1}}\leq \sum_{j=0}^{N-1} (2^{\frac{\alpha}{2}}-1)^{w_H(\vec{j})}{\tau_a^\alpha}_{AB_j}
\end{eqnarray}
for $0\leq\alpha\leq2$, where $\vec{j}=(j_0,j_1,\cdots,j_{N-1})$ is the vector from the binary representation of $j$ and $w_H(\vec{j})$ is the Hamming weight of $\vec{j}$.

{\sf [Proof].} Without loss of generality, we can always have
\begin{eqnarray}\label{tau11}
{\tau_a}_{AB_j}\geq {\tau_a}_{AB_{j+1}}\geq 0,
\end{eqnarray}
by relabeling the subsystems.
From (\ref{cb}), it is sufficient to show that
\begin{eqnarray}\label{pfth11}
\left(\sum_{j=0}^{N-1} {\tau_a^2}_{AB_j}\right)^\frac{\alpha}{2}\leq \sum_{j=0}^{N-1} (2^{\frac{\alpha}{2}}-1)^{w_H(\vec{j})}{\tau_a^\alpha}_{AB_j}.
\end{eqnarray}
We first prove the inequality (\ref{pfth11}) for the case that $N$ is a power of 2, $N = 2^n$, by mathematical induction.
For $n = 1$, by using Lemma 1 we have
\begin{equation*}\label{}
  {\tau_a^\alpha}_{A|B_0B_1}\leq  {\tau_a^\alpha}_{AB_0}+(2^{\frac{\alpha}{2}}-1){\tau_a^\alpha}_{AB_1},
\end{equation*}
which is just the inequality (\ref{pfth11}) for $N=2$.

Now let us assume that the inequality (\ref{pfth11}) is true for $N=2^{n-1}$ with $n \geq 2$, and consider the case that $N=2^n$.
For an $(N+1)$-partite quantum state $\rho_{AB_0\cdots B_{N-1}}$ and its bipartite reduced density matrices $\rho_{AB_j}$, $j=0,1,\cdots, N-1$, we have
\begin{eqnarray}\label{pfth12}
&&\left(\sum_{j=0}^{N-1} {\tau_a^2}_{AB_j}\right)^\frac{\alpha}{2}\nonumber\\
&&= \left(\sum_{j=0}^{2^{n-1}-1} {\tau_a^2}_{AB_j}+\sum_{j=2^{n-1}}^{2^n-1} {\tau_a^2}_{AB_j}\right)^\frac{\alpha}{2}\nonumber\\
&&= \left(\sum_{j=0}^{2^{n-1}-1} {\tau_a^2}_{AB_j}\right)^\frac{\alpha}{2}\left(1+\frac{\sum_{j=2^{n-1}}^{2^n-1} {\tau_a^2}_{AB_j}}{\sum_{j=0}^{2^{n-1}-1} {\tau_a^2}_{AB_j}}\right)^\frac{\alpha}{2}.
\end{eqnarray}

Due to (\ref{tau11}) we have
\begin{eqnarray}\label{pfth13}
\sum_{j=2^{n-1}}^{2^n-1} {\tau_a^2}_{AB_j}\leq\sum_{j=0}^{2^{n-1}-1} {\tau_a^2}_{AB_j}.
\end{eqnarray}
By using Lemma 1 we get
\begin{eqnarray}\label{pfth14}
\left(\displaystyle\sum_{j=0}^{N-1} {\tau_a^2}_{AB_j}\right)^\frac{\alpha}{2} & \leq\left(\displaystyle\sum_{j=0}^{2^{n-1}-1} {\tau_a^2}_{AB_j}\right)^\frac{\alpha}{2}\nonumber\\
& +(2^{\frac{\alpha}{2}}-1)\left(\displaystyle\sum_{j=2^{n-1}}^{2^n-1} {\tau_a^2}_{AB_j}\right)^\frac{\alpha}{2}.
\end{eqnarray}
Here, the induction hypothesis assures that
\begin{eqnarray}\label{pfth15}
\left(\sum_{j=0}^{2^{n-1}-1} {\tau_a^2}_{AB_j}\right)^\frac{\alpha}{2}\leq \sum_{j=0}^{2^{n-1}-1} (2^{\frac{\alpha}{2}}-1)^{w_H(\vec{j})}{\tau_a^\alpha}_{AB_j}.
\end{eqnarray}
From above relations we obtain
\begin{eqnarray}\label{pfth16}
\left(\sum_{j=2^{n-1}}^{2^{n}-1} {\tau_a^2}_{AB_j}\right)^\frac{\alpha}{2}\leq \sum_{j=2^{n-1}}^{2^{n}-1} (2^{\frac{\alpha}{2}}-1)^{w_H(\vec{j})-1}{\tau_a^\alpha}_{AB_j}.
\end{eqnarray}
Taking into account (\ref{pfth14}), (\ref{pfth15}) and (\ref{pfth16}) we have
\begin{eqnarray}\label{pfth17}
\left(\sum_{j=0}^{2^{n}-1} {\tau_a^2}_{AB_j}\right)^\frac{\alpha}{2}\leq \sum_{j=0}^{2^{n}-1} (2^{\frac{\alpha}{2}}-1)^{w_H(\vec{j})}{\tau_a^\alpha}_{AB_j} ,
\end{eqnarray}
which proves the inequality (\ref{pfth11}) for $N=2^n$.

Now for an arbitrary positive integer $N$, consider an $(N + 1)$-partite state $\rho_{AB_0\cdots B_{N-1}}$.
We can always assume that $0\leq N\leq 2^n$ for some $n$. Consider a $(2^n+1)$-partite quantum state
\begin{eqnarray}\label{pfth18}
\rho'_{AB_0\cdots B_{2^n-1}}=\rho_{AB_0\cdots B_{N-1}}\otimes \delta_{B_N\cdots B_{2^n-1}},
\end{eqnarray}
which is a product of $\rho_{AB_0\cdots B_{N-1}}$ and an arbitrary $(2^n -N )$-partite
quantum state $\delta_{B_N\cdots B_{2^n-1}}$.

Because $\rho'_{AB_0\cdots B_{2^n-1}}$ is a $(2^n +1)$-partite state, inequality (\ref{pfth17}) leads to
\begin{eqnarray}\label{pfth19}
{\tau_a^\alpha}(\rho'_{A|B_0B_1\cdots B_{2^n-1}})\leq \sum_{j=0}^{2^n-1} (2^{\frac{\alpha}{2}}-1)^{w_H(\vec{j})}{\tau_a^\alpha}(\sigma_{AB_j}),
\end{eqnarray}
where $\sigma_{AB_j}$ is the bipartite reduced density matrix of $\rho'_{AB_0\cdots B_{2^n-1}}$ for $j=0,1,\cdots,2^n-1$.
Since $\rho'_{AB_0\cdots B_{2^n-1}}$ is a separable state with respect to the bipartition between $AB_0\cdots B_{N-1}$ and $B_N\cdots B_{2^n-1}$, one has
\begin{eqnarray}\label{pfth110}
{\tau_a}(\rho'_{A|B_0B_1\cdots B_{2^n-1}})={\tau_a}(\rho_{A|B_0B_1\cdots B_{N-1}}),
\end{eqnarray}
and
\begin{eqnarray}\label{pfth111}
{\tau_a}(\sigma_{AB_j})=0,
\end{eqnarray}
for $j=N,\cdots,2^n-1$. Moreover, for $j=0,1,\cdots,N-1$ one has
\begin{eqnarray}\label{pfth112}
\sigma_{AB_j}=\rho_{AB_j}.
\end{eqnarray}

From (\ref{pfth19} - \ref{pfth112}), we have
\begin{eqnarray}\label{pfth113}
&&{\tau_a^\alpha}(\rho_{A|B_0B_1\cdots B_{N-1}})\nonumber\\
&&={\tau_a^\alpha}(\rho'_{A|B_0B_1\cdots B_{2^n-1}})\nonumber\\
&&\leq \sum_{j=0}^{2^n-1} (2^{\frac{\alpha}{2}}-1)^{w_H(\vec{j})}{\tau_a^\alpha}(\sigma_{AB_j})\nonumber\\
&&=\sum_{j=0}^{N-1} (2^{\frac{\alpha}{2}}-1)^{w_H(\vec{j})}{\tau_a^\alpha}(\rho_{AB_j}).
\end{eqnarray}
This completes the proof. $\Box$

We have obtained the general polygamy inequality of the $\alpha$th $(0\leq \alpha\leq 2)$ power of concurrence of assistance for arbitrary-dimensional quantum systems. In fact, (\ref{cb}) is a special case of (\ref{th12}) for $\alpha=2$. Besides, based on the improved polygamy relations, we get a new upper bound for bipartite entanglement in multipartite systems for $0\leq \alpha < 2$, which is better than (\ref{cb}). To illustrate the advantage of (\ref{th12}), we give an example as follows.

Let us consider the three-qubit state $\rho=|\psi\rangle\langle\psi|$ in the generalized Schmidt decomposition form,
where $|\psi\rangle=\lambda_0|000\rangle+\lambda_1e^{i{\varphi}}|100\rangle+\lambda_2|101\rangle+\lambda_3|110\rangle+\lambda_4|111\rangle$, $\lambda_i\geq0,~i=0,1,2,3,4$ and $\sum\limits_{i=0}\limits^4\lambda_i^2=1.$ We have
${\tau_a}_{A|BC}=2\lambda_0\sqrt{{\lambda_2^2+\lambda_3^2+\lambda_4^2}},$
${\tau_a}_{AB}=2\lambda_0\sqrt{{\lambda_2^2+\lambda_4^2}}$, and ${\tau_a}_{AC}=2\lambda_0\sqrt{{\lambda_3^2+\lambda_4^2}}$.
Take $\lambda_{0}=\lambda_{1}=\frac{1}{2}$, $\lambda_{2}=\lambda_{3}=\lambda_{4}=\frac{\sqrt{6}}{6}$,
one has ${\tau_a}_{A|BC}=\frac{\sqrt{2}}{2}$, ${\tau_a}_{AB}={\tau_a}_{AC}=\frac{\sqrt{3}}{3}$, and the marginal quantum relations
is ${\tau_a^2}_{AB}+{\tau_a^2}_{AC}-{\tau_a^2}_{A|BC}\approx0.167$ for $\alpha=2$.
For $\alpha=1$, the marginal quantum relations from (\ref{th12}) is ${\tau_a}_{AB}+ (\sqrt{2}-1){\tau_a}_{AC}-{\tau_a}_{A|BC}\approx0.109$,
which is smaller than the one for $\alpha=2$.

Since $0\leq(2^{\frac{\alpha}{2}}-1)^{w_H(\vec{j})}\leq 1$ for any $0\leq\alpha\leq 2$, we have
\begin{eqnarray}\label{pfth114}
{\tau_a^\alpha}_{A|B_0B_1\cdots B_{N-1}}&&\leq\sum_{j=0}^{N-1} (2^{\frac{\alpha}{2}}-1)^{w_H(\vec{j})}{\tau_a^\alpha}_{AB_j}\nonumber\\
&&\leq\sum_{j=0}^{N-1} {\tau_a^\alpha}_{AB_j},
\end{eqnarray}
for any multipartite quantum state $\rho_{A|B_0B_1\cdots B_{N-1}}$. Thus, we have the following corollary.

{[\bf Corollary 1]}. For any multiparty pure state $\rho_{AB_0\cdots B_{N-1}}$, we have
\begin{eqnarray*}\label{C12}
{\tau_a}^\alpha_{A|B_0B_1\cdots B_{N-1}}\leq \sum_{j=0}^{N-1}{\tau_a}^\alpha_{AB_j}
\end{eqnarray*}
for $0\leq\alpha\leq2$.

The class of weighted polygamy inequalities in Theorem 1 can be further tightened under some condition on bipartite quantum relations.

{\bf [Theorem 2]}.  For any multipartite pure state $\rho_{AB_0\cdots B_{N-1}}$, if
\begin{eqnarray}\label{th22}
{\tau_a^2}_{AB_i}\geq\sum_{j=i+1}^{N-1} {\tau_a^2}_{AB_j}
\end{eqnarray}
for $i=0,1,\cdots N-2$, we have
\begin{eqnarray}\label{th21}
{\tau_a^\alpha}_{A|B_0B_1\cdots B_{N-1}}\leq\sum_{j=0}^{N-1} (2^{\frac{\alpha}{2}}-1)^j{\tau_a^\alpha}_{AB_j}
\end{eqnarray}
for $0\leq \alpha\leq 2$.

{\sf [Proof].}
From Lemma 1, we have
\begin{eqnarray*}\label{pfth21}
&&{\tau_a^\alpha}_{A|B_0B_1\cdots B_{N-1}}\nonumber\\
&&\leq  {\tau_a^\alpha}_{AB_0}+(2^{\frac{\alpha}{2}}-1) \left(\sum_{j=1}^{N-1}{\tau_a^2}_{AB_j}\right)^\frac{\alpha}{2}\nonumber\\
&&\leq {\tau_a^\alpha}_{AB_0}+(2^{\frac{\alpha}{2}}-1){\tau_a^\alpha}_{AB_1}
 +(2^{\frac{\alpha}{2}}-1)^2 \left(\sum_{j=2}^{N-1}{\tau_a^2}_{AB_j}\right)^\frac{\alpha}{2}\nonumber\\
&& \leq \cdots\nonumber\\
&&\leq {\tau_a^\alpha}_{AB_0}+(2^{\frac{\alpha}{2}}-1){\tau_a^\alpha}_{AB_1}+\cdots+(2^{\frac{\alpha}{2}}-1)^{N-1}{\tau_a^\alpha}_{AB_{N-1}}.
\end{eqnarray*}
$\Box$

In Theorem 2, the condition (\ref{th22}) are always satisfied by some states. 
Let us consider a four-qubit state $\rho=|W\rangle_{ABCD}\langle W|$, where
$|W\rangle_{ABCD}=a|1000\rangle+b|0100\rangle+c|0010\rangle+d|0001\rangle$, and $a^2+b^2+c^2+d^2=1$. We have $\tau_a(\rho_{A|BCD})=2a\sqrt{1-a^2}$, $\tau_a(\rho_{AB})=2ab$, $\tau_a(\rho_{AC})=2ac$, $\tau_a(\rho_{AD})=2ad$. The condition (\ref{th22}) is satisfied as long as $b^2\geq c^2+d^2$. For example, we set $b=\frac{1}{\sqrt{2}}$, $a=c=d=\frac{1}{\sqrt{6}}$. Then the state $\rho=|W\rangle_{ABCD}\langle W|$ satisfies the condition (\ref{th22}). On the other hand, if $b^2\leq c^2+d^2$, e.g., 
$c=\frac{1}{\sqrt{2}}$ and $a=b=d=\frac{1}{\sqrt{6}}$, then $\rho$ does not satisfy the condition (\ref{th22}).

{\sf [Remark 1].}
For any non-negative integer $j$ and the corresponding binary vector $\vec{j}$ in Eq. (\ref{wj}), the Hamming weight $w_H(\vec{j})$ is upper bounded by $\log_2j$. Thus, we have
$w_H(\vec{j})\leq \log_2j\leq j$, which implies that
${\tau_a^\alpha}_{A|B_0B_1\cdots B_{N-1}}\leq\sum_{j=0}^{N-1} (2^{\frac{\alpha}{2}}-1)^j{\tau_a^\alpha}_{AB_j}\leq\sum_{j=0}^{N-1} (2^{\frac{\alpha}{2}}-1)^{w_H(\vec{j})}{\tau_a^\alpha}_{AB_j}$,
for $0\leq\alpha\leq 2$. In other words, inequality (\ref{th21}) in Theorem 2 is tighter than the inequality (\ref{th12}) in Theorem 1 for states satisfying the conditions ${\tau_a^2}_{AB_i}\geq\sum_{j=i+1}^{N-1} {\tau_a^2}_{AB_j}$, $i=0,\cdots,N-2$.

\section{POLYGAMY RELATIONS FOR entanglement of assistance}

Now we study the polygamy relations for entanglement of assistance.
For a tripartite pure state $|\psi\rangle_{ABC}$, one gets \cite{fgj} $E(|\psi\rangle_{A|BC})\leq E_a(\rho_{AB})+E_a(\rho_{AC})$,
where $E(|\psi\rangle_{A|BC})=S(\rho_A)$ is the entanglement between $A$ and $BC$ based on the von Neumann entropy
$S(\rho)=-\mathrm{Tr}\rho\ln\rho$, and $E_a(\rho_{AB})=\max\sum_ip_iE(|\psi_i\rangle_{AB})$, with the maximization taking over all possible pure state
decompositions of $\rho_{AB}=\sum_ip_i|\psi_i\rangle_{AB}\langle\psi_i|$. Later, in Ref. \cite{062302}, the authors presented a general polygamy relation
for any multipartite state $\rho_{A_1|A_2\cdots A_n}$,
\begin{eqnarray}\label{ea}
E_a(\rho_{A_1|A_2\cdots A_n})\leq \sum_{i=2}^nE_a(\rho_{A_1A_i}).
\end{eqnarray}

Recently, in Ref.\cite{042332}, the authors presented another class of multipartite polygamy inequalities based on the $\beta$th power of entanglement of assistance (EOA).
For any multipartite state $\rho_{A|B_0B_1\cdots B_{N-1}}$ and $0\leq\beta\leq 1$,
\begin{eqnarray}\label{e1}
{E_a^\beta}_{A|B_0B_1\cdots B_{N-1}}\leq\sum_{j=0}^{N-1} \beta^{w_H(\vec{j})}{E_a^\beta}_{AB_j},
\end{eqnarray}
if ${E_a}_{AB_i}\geq {E_a}_{AB_{i+1}}$ for $i=0,1,\cdots,N-2$; and
\begin{eqnarray*}\label{e2}
{E_a^\beta}_{A|B_0B_1\cdots B_{N-1}}\leq\sum_{j=0}^{N-1} \beta^j{E_a^\beta}_{AB_j},
\end{eqnarray*}
if ${E_a}_{AB_i}\geq \sum_{j=i+1}^{N-1}{E_a}_{A|B_j}$ for $i=0,1,\cdots,N-2$.
With a similar consideration to $\tau_{AB_0\cdots B_{N-1}}$, we have the following result for EOA.

{[\bf Theorem 3]}. For any multipartite state $\rho_{AB_0\cdots B_{N-1}}$, we have
\begin{eqnarray}\label{th3}
{E_a^\beta}_{A|B_0B_1\cdots B_{N-1}}\leq \sum_{j=0}^{N-1} (2^\beta-1)^{w_H(\vec{j})}{E_a^\beta}_{AB_j}
\end{eqnarray}
for $0\leq\beta\leq1$.

To illustrate the tightness of the inequality (\ref{th3}) compared with the  inequality (\ref{e1}) in \cite{042332}, we consider the three-qubit state $\rho_{ABC}=|W\rangle_{ABC}\langle W|$, where
$|W\rangle_{ABC}=\frac{1}{\sqrt{3}}(|100\rangle+|010\rangle+|001\rangle)$. We have $E_a(\rho_{A|BC})=S(\rho_A)=\log_23-\frac{2}{3}$ and
$E_a(\rho_{AB})=E_a(\rho_{AC})=\frac{2}{3}$. Thus the marginal EOA from inequality (\ref{e1}) is
${E_a^\beta}(\rho_{AB})+\beta{E_a^\beta}(\rho_{AC})-E_a(\rho_{A|BC})=(1+\beta)(\frac{2}{3})^\beta+\frac{2}{3}-\log_23$.
The marginal EOA from inequality (\ref{th3}) is ${E_a^\beta}(\rho_{AB})+(2^\beta-1){E_a^\beta}(\rho_{AC})-E_a(\rho_{A|BC})=2^\beta(\frac{2}{3})^\beta+\frac{2}{3}-\log_23$.
Fig. 1 shows that our inequality gives a smaller upper bound than (\ref{e1}) in \cite{042332}, namely, our marginal EOA is smaller than inequallity (\ref{e1}) in \cite{042332}
for $0<\beta<1$.

\begin{figure}
  \centering
  \includegraphics[width=8cm]{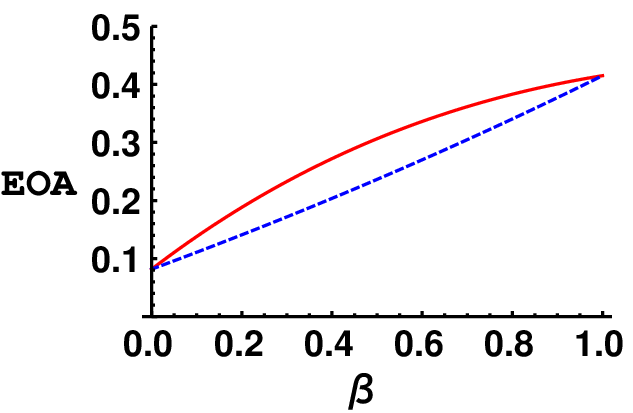}\\
  \caption{The blue dashed line represents the marginal EOA from inequality (\ref{th3}) for three-qubit $W$ state,
  the red thick line represents the marginal EOA from inequality (\ref{e1}) in \cite{042332}.}\label{2}
\end{figure}

Since $0\leq(2^\beta-1)^{w_H(\vec{j})}\leq 1$ for any $0\leq\beta\leq 1$, we have
\begin{eqnarray*}\label{}
{E_a^\beta}_{A|B_0B_1\cdots B_{N-1}}&&\leq\sum_{j=0}^{N-1} (2^\beta-1)^{w_H(\vec{j})}{E_a^\beta}_{AB_j}\nonumber\\
&&\leq\sum_{j=0}^{N-1} {E_a^\beta}_{AB_j},
\end{eqnarray*}
for any multipartite quantum state $\rho_{A|B_0B_1\cdots B_{N-1}}$. Thus, we have the following corollary.

{[\bf Corollary 2]}. For any multipartite pure state $\rho_{AB_0\cdots B_{N-1}}$, we have
\begin{eqnarray*}\label{C12}
{E_a}^\beta_{A|B_0B_1\cdots B_{N-1}}\leq \sum_{j=0}^{N-1}{E_a}^\beta_{AB_j}
\end{eqnarray*}
for $0\leq\beta\leq1$.

With a similar consideration to Theorem 2, we can tighten the class of weighted polygamy inequalities in Theorem 3 under certain conditions on bipartite quantum correlations.

{\bf [Theorem 4]}. For any multipartite state $\rho_{AB_0\cdots B_{N-1}}$, we have
\begin{eqnarray}\label{th4}
{E_a^\beta}_{A|B_0B_1\cdots B_{N-1}}\leq\sum_{j=0}^{N-1} (2^\beta-1)^j{E_a^\beta}_{AB_j},
\end{eqnarray}
conditioned that
\begin{eqnarray}\label{th42}
{E_a^2}_{AB_i}\geq\sum_{j=i+1}^{N-1} {E_a^2}_{AB_j},
\end{eqnarray}
for $i=0,1,\cdots N-2$, $0\leq \beta\leq 1$.

{\sf [Remark 2].} For any non-negative integer $j$, since $w_H(\vec{j})\leq \log_2j\leq j$, one has
${E_a^\beta}_{A|B_0B_1\cdots B_{N-1}}\leq\sum_{j=0}^{N-1} (2^\beta-1)^j{E_a^\beta}_{AB_j}\leq\sum_{j=0}^{N-1} (2^\beta-1)^{w_H(\vec{j})}{E_a^\beta}_{AB_j}$
for $0\leq\beta\leq 1$. Therefore, inequality (\ref{th4}) in Theorem 4 is tighter than the inequality (\ref{th3}) in Theorem 3 for
states satisfying the conditions ${E_a^2}_{AB_i}\geq\sum_{j=i+1}^{N-1} {E_a^2}_{AB_j}$, $i=0,\cdots,N-2$.

In particular, (\ref{th4}) reduces to (\ref{ea}) in \cite{062302} for $\beta=1$. For $0<\beta<1$, (\ref{th4}) is a tighter polygamy inequality compared with (\ref{ea}). Since $w_H(\vec{j})\leq j$, (\ref{th4}) in Theorem 4 is in general tighter than the (\ref{th3}) in Theorem 3. From the example shown in Fig. 1, one can see that (\ref{th3}) is generally tighter than the result in \cite{042332}. Hence our weighted polygamy relations give finer characterizations of the entanglement distributions among the subsystems, and help better security analysis of quantum key distribution \cite{MP} in quantum information processing.

\section{conclusion}

We have investigated the polygamy relations related to the concurrence of assistance. General polygamy inequalities given by the $\alpha$th $(0\leq \alpha\leq 2)$ power of concurrence of assistance have been presented for multipartite states in arbitrary-dimensional quantum systems. We have further shown that the general polygamy inequalities can even be improved to be tighter ones under certain conditions on the assisted entanglement of bipartite subsystems. Based on the improved polygamy relations, lower bound for distribution of bipartite entanglement has been provided for multipartite systems. Moreover, the $\beta$th ($0\leq \beta \leq 1$) power of polygamy inequalities have been obtained for the entanglement of assistance as a by-product, which are shown to be tighter than the existing ones.

The higher-dimensional quantum systems are the key resources in various quantum information and communication processing tasks.
For instance, the qudit ($d>2$) systems are preferred in some quantum key distributions, where the use of qudits increases
the coding density and provides stronger security compared to qubits \cite{sta}.
Our results apply to general polygamy relations of multipartite entanglement in arbitrary higher-dimensional quantum systems.
Moreover, our polygamy inequalities provide tighter constraints and finer characterizations of the entanglement distributions
among the multipartite systems. These results may highlight future works on the study of multipartite quantum entanglement.
\bigskip

\noindent{\bf Acknowledgments}\, \, This work is supported by the NSF of China under Grant No. 11847209; 11675113  and Key Project of Beijing Municipal Commission of Education under No. KZ201810028042.

\end{document}